\documentclass[11pt]{article}
\pdfoutput=1 
\usepackage{geometry}                
\usepackage{graphicx}
\usepackage{amssymb}
\usepackage{amsmath}
\usepackage{amsfonts}
\usepackage{graphicx}
\usepackage{epstopdf}
\usepackage{footmisc}
\usepackage{color}
\usepackage{cite}
\usepackage{soul}
\newcommand{\be}{\begin{equation}}
\newcommand{\ee}{\end{equation}}
\newcommand{\bea}{\begin{eqnarray}}
\newcommand{\eea}{\end{eqnarray}}
\newcommand{\beas}{\begin{eqnarray*}}
\newcommand{\eeas}{\end{eqnarray*}}
\newcommand{\w}{\vec{w}}
\newcommand{\x} {{\vec x}}

\title{Analytic solution to variance optimization with no short-selling}
\author{Imre Kondor$^{1,2,3}$, G\'abor Papp$^4$, Fabio Caccioli$^{5,6}$\\
{\it 1-Parmenides Foundation, Pullach, Germany}\\
{\it 2- Department of Investment and Corporate Finance, Corvinus University of Budapest,} \\
{\it Budapest, Hungary}
{\it 3- London Mathematical Laboratory, London, UK} \\
{\it 4- E\"otv\"os Lor\'and University, Institute for Physics, Budapest, Hungary } \\
{\it 5- University College London, Department of Computer Science,} \\
{\it London, WC1E 6BT, UK} \\
{\it 6- Systemic Risk Centre, London School of Economics and Political Sciences, London, UK}\\
}

\numberwithin{equation}{section}
\begin{document}
\maketitle

\abstract{A portfolio of independent, but not identically distributed, returns is optimized under the variance risk measure, in the high-dimensional limit where the number $N$ of the different assets in the portfolio and the sample size $T$ are assumed large with their ratio $r=N/T$ kept finite, with a ban on short positions. To the best of our knowledge, this is the first time such a constrained optimization is carried out analytically, which is made possible by the application of methods borrowed from the theory of disordered systems. The no-short selling constraint acts as an asymmetric $\ell_1$ regularizer, setting some of the portfolio weights to zero and keeping the out of sample estimator for the variance bounded, avoiding the divergence present in the non-regularized case. However, the susceptibility, i.e. the sensitivity of the optimal portfolio weights to changes in the returns, diverges at a critical value $r=2$. This means that a ban on short positions does not prevent the phase transition in the optimization problem, it merely shifts the critical point from its non-regularized value of $r=1$ to $2$. At $r=2$ the out of sample estimator for the portfolio variance stays finite and the estimated in-sample variance vanishes. 
We have performed numerical simulations to support the analytic results and found perfect agreement for $N/T<2$. Numerical experiments on finite size samples of symmetrically distributed returns show that above this critical point the probability of finding solutions with zero in-sample variance increases rapidly with increasing $N$, becoming one in the large $N$ limit. However, these are not legitimate solutions of the optimization problem, as they are infinitely sensitive to any change in the input parameters, in particular they will wildly fluctuate from sample to sample. With some narrative license we may say that the regularizer takes care of the longitudinal fluctuations of the optimal weight vector, but does not eliminate the divergent transverse fluctuations.
We also calculate the distribution of the optimal weights over the random samples and show that the regularizer preferentially removes the assets with large variances, in accord with one's natural expectation. 
}

\section{Introduction}
Institutional portfolios are often optimized under a ban on short positions. If the distribution of the returns on the securities making up the portfolio is exactly known, the optimization is straightforward to carry out. In practice, this distribution is never known, but has to be inferred from observations in the market. If the available data is finite, the optimal estimated portfolio weights will be different from their true values, and the resulting portfolio will suffer from estimation error. This error will be particularly large if the dimension $N$ of the portfolio (the number of different assets) is not small relative to the sample size (the length of available time series) $T$.  
This problem has been approached by various numerical methods, see e.g. \cite{scherermartin} for an overview.  In real life context of risk management or asset management a purely numerical approach may, however, be very computationally demanding and, as will be discussed below, may in addition be also misleading, especially if one lacks a full control over the optimization algorithm implemented in the risk management package and a good understanding of the structure of the problem. 

Such an understanding can come from an analytic approach. Analytic calculations of the optimal estimated portfolio have been performed by various groups under the assumption that the underlying statistical distribution is normal, the objective function is the variance and the optimization is subject to the budget constraint and, in some cases, an $\ell_2$ regularizer \cite{jobson1979Improved,jorion1986Bayes,Jagannathan2003,ledoit2003Improved,ledoit2004Honey,ledoit2004AWell,kempf2006Estimating,         Okhrin2006Distributional,golosnoy2007Multivariate,frahm2008,Basak2009Jackknife,DeMiguel2009,DeMiguel2009Optimal,frahm2010Dominating}. The most recent, nonlinear realization of $\ell_2$ shrinkage \cite{ledoit2011eigenvectors,ledoit2012nonlinear,Bun2016Laundrette} has turned out to be particularly effective in suppressing sample fluctuations. A special approach to portfolio optimization \cite{Ciliberti2007On,ciliberti2007Risk,caccioli2013Optimal,caccioli2015Portfolio,kondor2015Contour,caccioli2016Lp,Papp2016Variance,Shinzato2016Replica} rests on the replica method borrowed from the statistical physics of disordered systems  \cite{mezard1987Spin}. These papers focused on the minimal risk portfolio, but \cite{varga2016Replica} treated the full Markowitz problem \cite{markowitz1952Portfolio} including the constraint on the expected return, while in \cite{Papp2016Variance,Shinzato2016Minimal} an $\ell_2$  constraint has been imposed on the portfolio weights. Such a regularizer can suppress large sample fluctuations that lead to a high degree of estimation error, especially in the high dimensional setting where both the dimension $N$ and the sample size $T$ are large. An alternative motivation for an $\ell_2$ constraint is to prevent the over-concentration of the optimal portfolio on a small number of blue chips  \cite{bouchaud2003Theory,gabor1999Portfolios,Shinzato2016Minimal}, a particularly strong tendency in small markets, and also by taking into account the market impact of a future liquidation of the portfolio already at the stage of its composition \cite{caccioli2013Optimal}.

Considerations of transaction costs and the technical difficulty of frequent rebalancing a very large portfolio may make it desirable to reduce the dimension and strive for a sparse portfolio. This can be achieved by borrowing the popular and very successful $\ell_1$ regularization from machine learning \cite{hastie2008Elements}. Jagannathan and Ma \cite{Jagannathan2003} were the first to notice that a ban on short positions, which can be regarded as a special case of  $\ell_1$ regularization, improves the stability of estimated optimal portfolios. Subsequently Brodie et al \cite{brodie2009Sparse} applied an $\ell_1$ regularizer on the portfolio weights in an empirical study of real life portfolios in various markets and demonstrated its satisfactory performance compared with the $1/N$ portfolio \cite{DeMiguel2009Optimal}. 

To the best of our knowledge, no analytic result exists in the literature for portfolio optimization under an $\ell_1$ constraint. The purpose of the present paper is to perform such an analytic optimization of the variance as the risk measure supplemented with a special case of the $\ell_1$ constraint, a ban on negative portfolio weights. The method that makes this possible is again the replica method. It assumes that return samples of size $T$ are drawn from an $N$-dimensional normal distribution with $N$ and $T$ going to infinity with their ratio $r=N/T$ kept finite. For simplicity, we will also assume that the expected return of each asset in the portfolio is zero and seek to determine the global minimal risk portfolio, but we allow the assets to have different variances. We are considering independent normal variables and assume that the returns are also serially independent (zero autocorrelation). 

We also analyze the numerical aspects of this problem and find that the simulations precisely follow the theoretical curves up to the critical point $N/T=2$. While our simulations clearly indicate that in the region of infinite susceptibility above this critical ratio there is no meaningful solution, numerical work above this point requires special care: some solvers (e.g. fmincon) modify the problem in order to make sure a stable solution exists even when the covariance matrix is less than full rank. Without a careful study of the algorithm's description and without anticipating the instability, it is easy to overlook the phase transition.

The plan of the rest of the paper is as follows. For the sake of establishing a basis for later comparison and introducing some notation, in Sec. 2 we address the trivial problem of optimizing the variance assuming we have complete information, as if having an infinitely large sample. In Sec. 3 we consider the case of variance optimization without the no-short constraint, but now for $r=N/T$ finite. Some of the results here reproduce those known previously, but the distribution of weights is new, as is also the discussion of the geometry of the phase transition (that in the unconstrained case takes place at $r=1$). Sec. 4 is the central part of the paper. Here, we perform the optimization of variance with a constraint forbidding short positions, and derive results for the estimator for the our of sample variance, the sensitivity to changes in the input data (susceptibility), and the in-sample estimator for the portfolio variance, along with results for the distribution of weights over the random samples. This constitutes a complete solution of the no-short constrained problem, the first instance such a solution has been achieved by analytic means. Our formulae illustrate how a ban on short selling removes some of the items from the portfolio, and how an asset's volatility affects the probability of its elimination. We identify the phase transition at $r=2$ mentioned above, which is different in nature from the one at $r=1$ in the unconstrained case in that the susceptibility diverges but the  estimation error stays finite here. Sec. 5 is a summary of the results. Technical details are relegated to two appendices. Appendix A presents the replica derivation of the free energy functional for the optimization of the variance supplemented by a generic constraint, while Appendix B derives the saddle point equations and the distribution of the weights.

\section{Optimizing the variance with complete information, $r = 0$}

In this section we present an analytic treatment of the optimization of the variance of a portfolio composed of $N$ securities with zero expected returns and a diagonal covariance matrix with given elements $\sigma^2_i$ along the diagonal,  $i=1,2,\ldots,N$. The risk $\sigma^2_p$ of the portfolio measured in terms of the variance is

\be\label{eq:Variance}
\sigma^2_p=\sum_i \sigma_i^2 w_i^2
\ee
to be minimized under the budget constraint

\be\label{eq:BudgetConstraint}
\sum_i w_i = N,
\ee
where, instead of the usual $1$, we normalized the portfolio weights $w_i$ to $N$, in order to keep them of order unity. (In the following we will consider the dimension $N$ of the portfolio as a large number, letting it go to infinity when the calculations so demand.) As the assets are assumed to have zero expected returns, we do not stipulate a constraint on the expected return of the portfolio, and seek the global minimum risk portfolio.

The optimization problem \eqref{eq:Variance}, \eqref{eq:BudgetConstraint} is trivial to solve by the method of Lagrange multipliers. The minimum of

\be\label{eq:Lagrangian}
\sum_i\sigma_i^2 w_i^2-\lambda(\sum_i w_i-N)
\ee
is at $w_i= \lambda/2\sigma_i^2$ , and the budget constraint fixes the Lagrange multiplier to be 

\be\label{eq:lagrangeMultiplier}
\lambda=\frac{2N}{\sum_i\frac{1}{\sigma_i^2}}.
\ee
The optimal portfolio weights are then obtained as

\be\label{eq:TrueWeights}
w_i^* = \frac{1}{\sigma_i^2}\frac{N}{\sum_j\frac{1}{\sigma_j^2}}
\ee
and the minimal risk is

\be\label{eq:TrueRisk}
{\sigma^*_p}^2=\frac{N}{\frac{1}{N}\sum_j\frac{1}{\sigma_j^2}}.
\ee
For later convenience we define 

\be\label{eq:FreeEnergy}
F=\frac{T{\sigma^*_p}^2}{2N}=\frac{1}{2 r}\frac{N}{\frac{1}{N}\sum_j\frac{1}{\sigma_j^2}} ,
\ee
and we will refer to this as the ``free energy'' or the cost function. The factor $1/(2r)$, where $r=N/T$, will then appear also in the Lagrange multiplier $\lambda$. If we define $\lambda'$ as the Lagrange multiplier associated with the minimization of the free energy \eqref{eq:FreeEnergy}, we have that
\be
\lambda'=\frac{1}{2r}\lambda=\frac{1}{2r}\frac{2}{\frac{1}{N}\sum_i\frac{1}{\sigma_i^2}}.
\ee
In the following we will always use $\lambda'$ everywhere, and will omit the prime with no risk of confusion.

Note that due to the normalization of the weights $F$ is of order $N$. In the following it will be convenient to consider the free energy per asset

\be\label{FreeEnergyper asset}
f=\frac{1}{2 r}\frac{1}{\frac{1}{N}\sum_j\frac{1}{\sigma_j^2}}=\frac{1}{2}\lambda
\ee

As already evident from \eqref{eq:Lagrangian}, the Lagrange multiplier associated with the budget constraint must be positive; a negative value would correspond to no security in the portfolio at all. Thus $\lambda$ plays a role analogous to the  chemical potential, the quantity that governs the number of particles in a physical system, and, for brevity, we will refer to $\lambda$ as the chemical potential in the following. The positivity of $\lambda$ is completely trivial at this point, but it will acquire significance in the computations later: its vanishing will herald the phase transition.

The optimal weights are the larger the smaller their variance, in particular, if one of the securities is riskless, its weight takes up the full weight $N$.
Also, if there is a riskless security in the portfolio, the whole portfolio becomes riskless and $\sigma^*_p$ vanishes.

Note also that the no-short selling condition did not have to be stipulated in this preliminary instance: the weights have worked out to be positive automatically. This will not remain true when the parameters of the model are estimated on the basis of finite samples.

The optimization problem as laid out above assumes that we have complete knowledge about the probability distribution of the returns: in particular we know the (zero) values of the expected returns and the values of the variances $\sigma_i$. In reality, we never have complete information. What we may have are samples of size $T$ drawn from the joint distribution of returns, which in our setting is
\be
P(\{x_{it}\})=\prod_i \left(\sqrt{\frac{N}{2\pi\sigma_i^2}}e^{-Nx_{it}^2/2\sigma_i^2}\right).
\ee
An important parameter of the problem is the ratio $r=N/T$. The larger the sample size $T$ relative to the dimension $N$, the better the estimates we can make for the optimal weights and the optimal value of the risk. We expect, therefore, that in the limit $r \to 0$ we can retrieve the ``true'' values of the weights as given in \eqref{eq:TrueWeights}, and the ``true'' value of the optimal risk, \eqref{eq:TrueRisk}.

Present day institutional portfolios are large, with $N$'s in the range of hundreds or thousands, while sample sizes are limited by stationarity considerations to below $1000$ (four years worth of daily data) at most, but often much less. Therefore, the value of $r$ is never really small in practice. This leads to large sample fluctuations, so large indeed that at a critical value of $r$ the estimation error becomes infinite and the optimization meaningless. In the case of unregularized variance as risk measure, this critical value is $r_c=1$, which is where the estimated covariance matrix loses its positive definiteness and the first zero eigenvalue appears.

Difficulties of a similar nature appear in countless problems in modern statistics and machine learning \cite{buhlmann2011statistics}. The remedy is to introduce regularizers, i.e.  terms added to the cost function with the purpose of suppressing the large sample fluctuations. Of course, regularization will also introduce bias, but the hope is that a reasonable balance can be struck between bias and fluctuations.

Perhaps the most popular regularizer today is the one based on the $\ell_1$ norm \cite{tibshirani1996regression}. Its appeal was greatly enhanced by the proof by Cand\`es et al. \cite{candes2006stable} that $\ell_1$ can successfully imitate $\ell_0$, the straight weeding out of the superfluous, irrelevant variables, thereby strongly reducing the dimension of the problem. In the portfolio context this would mean reducing the dimensionality by setting the weights to zero of the securities that are deemed irrelevant, presumably those with the largest volatilities.
In the following, we are going to demonstrate the action of $\ell_1$ regularization in the special case corresponding to a no-short selling constraint. Before addressing that problem, however, we wish to present the optimization of variance without the no-short constraint.

\section{Unconstrained variance optimization}
By ``unconstrained'' we mean dropping the no-short condition; the budget constraint will of course be upheld.

The relevant free energy functional is obtained from \eqref{eq:AppendixBFreeEnergy2} by setting $\eta_1=\eta_2=0$ and making use of the identity
\be
W(x)+W(-x)=\frac{x^2+1}{2},
\ee
satisfied by the transcendental function $W$ appearing in \eqref{eq:AppendixBFreeEnergy2}. Then $f$ works out to be
\be\label{eq:FUnconstrained}
f=\lambda-\Delta\hat q_0-\hat\Delta q_0+\frac{1}{2r} \frac{q_0}{1+\Delta}+\frac{\hat q_0}{2 \hat\Delta}-\frac{\lambda^2}{4\hat\Delta}\frac{1}{N}\sum_i\frac{1}{\sigma_i^2}.
\ee
Setting the derivatives of $f$ with respect to the ``order parameters'' $\lambda$, $q_0$, $\Delta$, $\hat\Delta$ and $\hat q_0$ to zero gives the following saddle-point or stationarity conditions:

\be\label{eq:SDUnconstrained1}
\lambda=2 \hat\Delta\left(\frac{1}{N}\sum_i\frac{1}{\sigma_i^2}\right)^{-1},
\ee

\be\label{eq:SDUnconstrained2}
\hat\Delta = \frac{1}{2 r}\frac{1}{1+\Delta},
\ee

\be\label{eq:SDUnconstrained3}
\hat q_0 = -\frac{q_0}{2 r (1+\Delta)^2},
\ee

\be\label{eq:SDUnconstrained4}
q_0 = -\frac{\hat q_0}{2\hat\Delta^2}+\frac{\lambda^2}{4 \hat\Delta^2}\frac{1}{N}\sum_i\frac{1}{\sigma_i^2},
\ee

\be\label{eq:SDUnconstrained5}
\Delta=\frac{1}{2\hat\Delta}.
\ee
Combining \eqref{eq:SDUnconstrained1}--\eqref{eq:SDUnconstrained5} one can easily see that the cost function $f$ at the saddle point is equal to

\be
f=\frac{\lambda}{2}.
\ee
The solution of the saddle point equations is straightforward:

\be\label{eq:SolutionSDUnconstrained1}
\lambda=\frac{1-r}{r}\frac{1}{\frac{1}{N}\sum_i\frac{1}{\sigma_i^2}},
\ee

\be\label{eq:SolutionSDUnconstrained2}
\Delta=\frac{r}{1-r},
\ee

\be\label{eq:SolutionSDUnconstrained3}
q_0 = \frac{1}{1-r}\frac{1}{\frac{1}{N}{\sum_i\frac{1}{\sigma_i^2}}},
\ee

\be\label{eq:SolutionSDUnconstrained4}
\hat q_0 = -\frac{1-r}{2r}\frac{1}{\frac{1}{N}{\sum_i\frac{1}{\sigma_i^2}}},
\ee

\be\label{eq:SolutionSDUnconstrained5}
\hat\Delta = \frac{1-r}{2r},
\ee
and the free energy per asset is
\be\label{eq:SolutionSDUnconstrainedF}
f = \frac{1-r}{2r}\frac{1}{\frac{1}{N}{\sum_i\frac{1}{\sigma_i^2}}}.
\ee
Turning to the distribution of weights, we see from \eqref{eq:AppendixBW1} and \eqref{eq:AppendixBW2} that for $\eta_1=\eta_2=0$    $w_i^{(1)}=w_i^{(2)}$, so the first term (the $\delta$-peak of the zero weights) in \eqref{eq:AppendixBWeightDistribution} vanishes, while the second term becomes

\be\label{eq:SolutionSDUnconstrainedWeightDistribution}
p(w) = \frac{1}{N}\sum_i\frac{1}{\sigma_w^{(i)}\sqrt{2\pi}}{\rm exp}\left(-\frac{1}{2}\left(\frac{w-w_0^{(i)}}{\sigma_w^{(i)}}\right)^2\right),
\ee
where

\be\label{eq:SolutionSDUnconstrainedWeights}
 w_0^{(i)}= \frac{\lambda}{2\sigma_i^2\hat\Delta} = \frac{\lambda r (1+\Delta)}{\sigma_i^2}
\ee
and

\be\label{eq:SolutionSDUnconstrainedSigmaW}
\sigma_w^{(i)} = \frac{\sqrt{q_0 r}}{\sigma_i} .
\ee

From \eqref{eq:SolutionSDUnconstrained1} and \eqref{eq:SolutionSDUnconstrained2} it follows that 

\be\label{eq:WeightsUnconstrained}
w_0^{(i)}= \frac{1}{\sigma_i^2}\frac{N}{\sum_j\frac{1}{\sigma_j^2}}  ,
\ee
the same as in \eqref{eq:TrueWeights}. Therefore, in the unconstrained optimization case the estimated weights fluctuate about their true values. This does not remain so once the no-short constraint is switched on.

The order parameters $\lambda$, $\Delta$ and $q_0$ have a direct meaning. As already seen in Section 2, $\lambda$ is the ``chemical potential'', the Lagrange multiplier associated with the budget constraint. As such, it must be positive, and its vanishing signals an instability. The quantity $\Delta$ was shown in \cite{caccioli2015Portfolio} to be the susceptibility, the measure of the sensitivity of the estimate to small changes in the input data. It is non-negative by definition, and its divergence is another signal of the instability that sets in for $\lambda=0$. Finally, $q_0$ is related to the out of sample estimator of the variance. In \cite{caccioli2015Portfolio} it was shown for the special case $\sigma_i=1,$ for all $i$, that $\sqrt{q_0}-1$ is the relative estimation error. When the variances of returns in the portfolio are different, $q_0$ has to be normalized as \cite{varga2016Replica}

\be\label{eqd:tildeQ}
\tilde q_0 = q_0\frac{1}{N}\sum_i\frac{1}{\sigma_i^2}=\frac{1}{1-r} ,
\ee
in order to make $\tilde q_0$ equal to the ratio between the optimal out of sample estimator for the risk of the portfolio (with weights $\hat w_i^*$ ) and the risk of the true optimal portfolio (with weights $w_i^*$)

\be
\tilde q_0 = \frac{\sum_{ij}\sigma_{ij} \hat w_i^*\hat w_j^*}{\sum_{ij}\sigma_{ij}  w_i^*  w_j^*}
\ee
so that $\sqrt {\tilde q_0}-1$ becomes the relative error associated with the estimation of risk. Because of the simple proportionality between $q_0$ and $\tilde q_0$, we will speak about $q_0$ as (the measure of) the out of sample estimation error. The divergence of $q_0$ or $\tilde q_0$ at $r=1$ is pointing to the same instability as that of $\Delta$ or the vanishing of $\lambda$.

\subsection{The limit of complete information}
When $r\to 0$, the sample size $T$ is much larger than the dimension $N$, so we have complete information and should be able to recover the results in Section 2.

This is indeed so: for $r\to0$ \eqref{eq:SolutionSDUnconstrained1} and \eqref{eq:SolutionSDUnconstrainedF} duly reproduce \eqref{eq:lagrangeMultiplier} and \eqref{eq:FreeEnergy}, respectively. From \eqref{eq:SolutionSDUnconstrained3} and \eqref{eqd:tildeQ} we also see that $\tilde q_0=1$, that is the estimation error vanishes.
Furthermore, \eqref{eq:SolutionSDUnconstrained2} implies that the susceptibility $\Delta$ vanishes with $r$. Then from \eqref{eq:SolutionSDUnconstrained1} and \eqref{eq:SolutionSDUnconstrainedWeights} it follows that

\be
w^{(i)}=\frac{1}{\sigma_i^2}\frac{1}{\frac{1}{N}\sum_j\frac{1}{\sigma_j^2}}
\ee
is the weight of asset $i$ in the optimal portfolio, in agreement with \eqref{eq:TrueWeights}.

The width $\sigma_w^{(i)}$ of the Gaussian distribution of the weights over the samples goes to zero with $r$, so the distribution \eqref{eq:SolutionSDUnconstrainedWeightDistribution} becomes a series of $\delta$-spikes
\be
p(w) = \frac{1}{N}\sum_i\delta\left(w-w^{(i)}\right),
\ee
where $\delta$ is the Dirac $\delta$-distribution.

\subsection{The high-dimensional case and the instability}

If $r$ is not very small, $N$ and $T$ become comparable and we are in the high-dimensional setting.
From \eqref{eq:SolutionSDUnconstrained1}-\eqref{eq:SolutionSDUnconstrained3} we see that with increasing $r$ the chemical potential $\lambda$ decreases, the susceptibility $\Delta$ increases, as does also the estimation error $q_0$, while the cost function $f$ decreases. As a result of averaging over the samples, the sharp peaks in the distribution of weights in \eqref{eq:SolutionSDUnconstrainedWeightDistribution} broaden.

As we approach $r=1$, the susceptibility $\Delta$ and the relative estimation error $q_0$ grow without bound, and the width of the Gaussian in \eqref{eq:SolutionSDUnconstrainedWeightDistribution} also diverges, so the different assets are not resolvable anymore. All these are signatures of an instability, divergent fluctuations from sample to sample, which we can rightly call a phase transition.

Note that in the same limit $r\to1$ the chemical potential $\lambda$ and the free energy $f$, the in-sample estimation of the cost, vanish.

The nature of this phase transition has been analyzed in detail in \cite{varga2016Replica}. At this point we merely point out that the replica method leading to the result in the present section can obviously not be continued beyond $r=1$, because as a method relying on a saddle point approximation (see Appendix A) it is bound to break down when the eigenvalues of the second derivatives of the replica functional all vanish, as it was demonstrated to be the case in \cite{varga2016Replica}.

On the other hand, there is nothing to prevent us from considering large dimensions and relatively small samples, that is a situation when $r>1$. What is happening in this region is the subject of the next subsection.

\subsection{Linear algebraic interpretation of the instability at $r=1$}

In the simple case of the variance, the root of the instability at $r=1$ is quite obvious; nevertheless it deserves a brief discussion here, especially because similar instabilities appear in several other risk measures including the Expected Shortfall \cite{Ciliberti2007On}, mean absolute deviation \cite{ciliberti2007Risk}, the minimax problem \cite{kondor2007Noise}, even in a GARCH-based non-stationary process \cite{varga2007Noise}, where they are considerably more difficult to explain. Moreover, we shall encounter a somewhat similar instability later when we introduce a constraint on short positions.

Let us consider the minimization of the empirical portfolio variance $\hat\sigma^2_p$ with the matrix of observed returns $x$. The empirical covariance matrix $C$ is given by

$$
C_{ij}=\frac{1}{T}\sum_t x_{it} x_{jt}
$$
and the empirical variance of the portfolio by

\be\label{eq:covmatrixassumofsquares}
\hat\sigma^2_p = \frac{1}{T}\sum_{ijt} w_i x_{it} x_{jt} w_j = \frac{1}{T}\sum_{t=1}^T\left(\sum_i w_i x_{it}\right)^2.
\ee
This is to be minimized over the weights $w_i$ subject to the budget constraint $\sum_i w_i=N$.

The rank of the covariance matrix $C$ is the smaller of $N$ and $T$ with probability one. The minimization of $\hat\sigma^2_p$ gives us $N$ equations which determine the solution as long as $N\le T$. When $N$ is larger than $T$, only $T$ of these equations are independent, so we have more unknowns than equations. For $N\ge T+1$ any weight vector selected from the null-space of $C$ will be a solution of the minimization problem, with $\hat\sigma^2_p=0$ as the minimal value of the cost function.

An alternative way to describe the situation is that with $N$ larger than $T$ the cost function will be flat along the directions lying in the null space of the covariance matrix and the solution can run away along these flat directions to an arbitrary distance from the origin. This means that arbitrarily large compensating positive and negative weights can arise, without violating the budget constraints and still keeping the portfolio variance at zero.

Arbitrarily large leverage combined with a zero value of the risk measure is a prescription for disaster. The first author to point out this dangerous feature of the variance was Jorion \cite{jorion1992portfolio}. A similar apparent arbitrage in Expected Shortfall and other downside risk measures was analyzed and identified as the root of instability in \cite{kondor2010instability,vargahaszonits2008instabilityofdownside,caccioli2013Optimal,caccioli2016Lp}.

It must be clear from the foregoing that this instability has nothing to do with the replica method, or the Gaussian distribution of returns, or the averaging over the samples. The root of this instability is purely geometrical, it arises in every single sample and for any underlying distribution of the returns, and it always takes place at the same critical ratio $r=N/T=1$. The universality of the critical value $r_c=1$ of the unconstrained variance optimization was demonstrated in \cite{varga2016Replica} and is a special case of the universality discussed by \cite{Donoho2009Observed} and \cite{amelunxen2013living}.

 \begin{center}
\begin{figure}[h]
\begin{center}
\includegraphics[width=10cm]{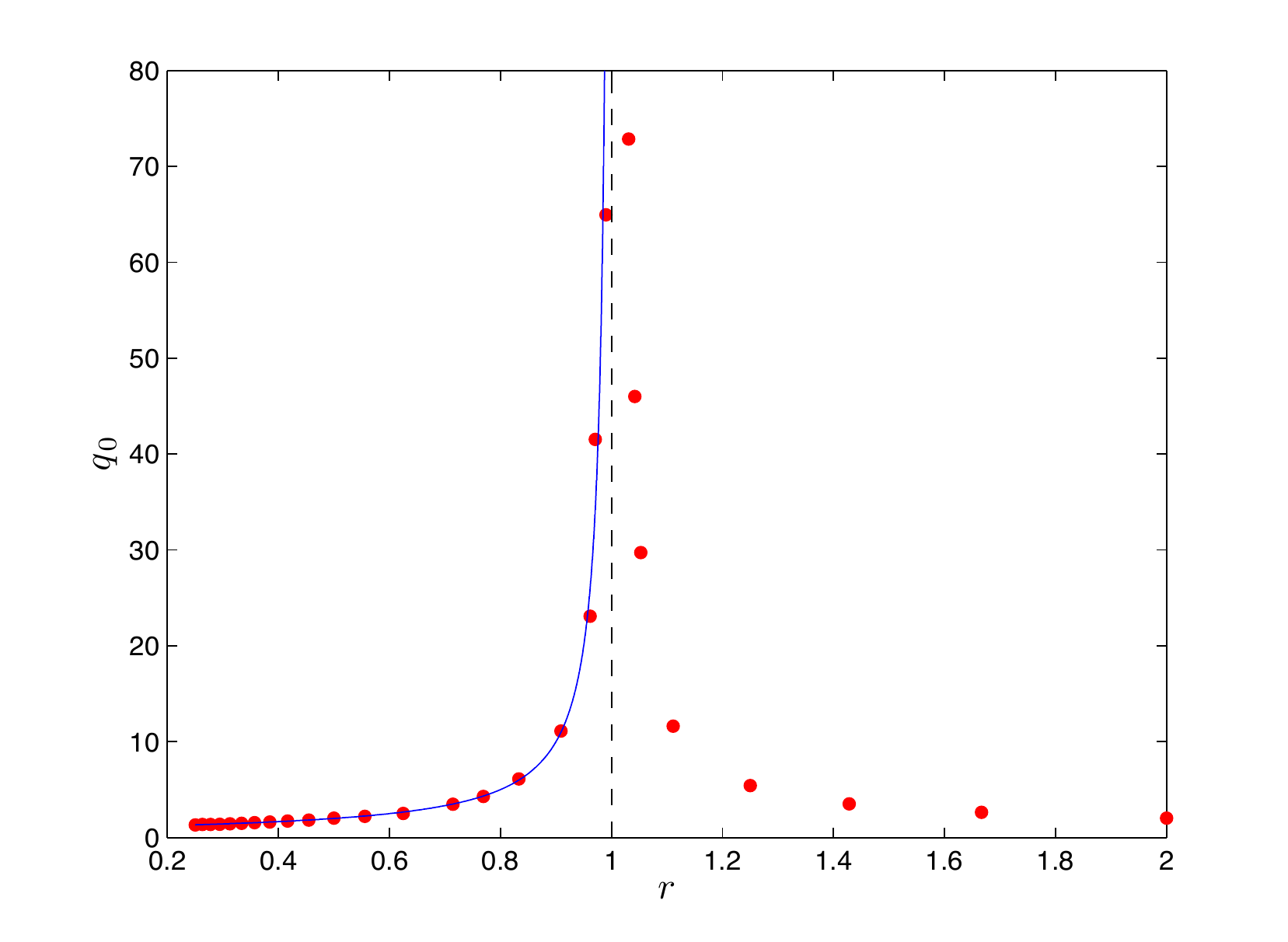}
\caption{\footnotesize {\textit{Estimation error as a function of $r$. The solid blue line represents the analytical solution obtained with the replica method. The dashed black line indicates the critical point $r_c=1$. Red dots represent results of numerical simulations averaged over $1000$ simulations for a system with $N=100$. Numerical simulations have been performed with Matlab using the function ``fmincon'' and the active-set algorithm. The match between numerical and analytical result is very good in the allowed region $r<1$. Due to a built-in regularizer in the solver, numerical solutions can be found also in the forbidden region. This could create the illusion that it is possible to find reliable solutions to the optimization problem also with very few data points.}}}
\end{center}
\label{chihom}
\end{figure}
\end{center}
To conclude this subsection, let us point out the significance of this instability for empirical work. Without additional constraints the instability must show up for any empirical sample with $N>T$. To check this, we generated synthetic time series of length $T$ for various values of $N$. For simplicity, we considered a set of assets with the same variance $\sigma_i=1$ for all $i$, and determined the optimal cost and the estimation error $q_0$ for $r$ values ranging from zero up to 2. The result of this numerical experiment performed with the Matlab solver ``fmincon'' is shown in Fig. 1. The surprising feature is that after a strong increase on approaching $r=1$, $q_0$ starts to decrease above $r=1$ again, as if the estimator became restabilized. Thus the program produces a stable result even in the region where we know that a continuum of equivalent solutions exist. The resolution of this puzzle lies in the fact that some of the numerically optimized solvers  contain what effectively amounts to an $\ell_2$ regularizer that does not influence the result as long as there is a meaningful one, but kicks in when a singular covariance matrix is encountered, and selects the diagonal vector ($w_i=1$ for all $i$) from among the infinitely many equivalent (and meaningless) solutions. Of course, this is properly indicated in the description of the solver, but easily overlooked by the user. This should be a warning to users against the blind application of ready-make programs without understanding their details and without a grasp of the main feature of the expected solution already before the numerical study.

The instability of the unconstrained variance has been pointed out several times earlier, and it is also easy to notice in empirical work from the ever-increasing sample fluctuations. This is not the case for the instability of the no-short-constrained variance optimization to which we turn now.

\section{Optimization with no short positions}

Portfolio optimization is often subject to constraints or an outright ban on short positions. Optimizing the variance under such conditions is a problem in quadratic programming that is routinely solved numerically. In this section we give what we believe to be the first analytic treatment of portfolio optimization with no short positions allowed.

The starting point is \eqref{eq:AppendixBFreeEnergy} and \eqref{eq:AppendixBPotential}. If we want to exclude negative weights, we impose infinite penalty on them by letting $\eta_2\to\infty$ in \eqref{eq:AppendixBPotential}. Positive positions will not be penalized, so we set $\eta_1=0$. According to \eqref{eq:AppendixBFreeEnergy2}--\eqref{eq:AppendixBSP5}, this leads to the free energy and stationarity conditions as follows:
\be\label{eq:constrainedFreeEnergy}
f=\lambda-\Delta \hat q_0-\hat\Delta q_0+\frac{1}{2r}\frac{q_0}{1+\Delta}+\frac{\hat q_0}{\hat\Delta}\frac{1}{N}\sum_i W\left(\frac{\lambda}{\sigma_i\sqrt{-2\hat q_0}}\right)
\ee

\be\label{eq:SPConstrained1}
\frac{1}{\sqrt{q_0 r}} = \frac{1}{N}\sum_i\frac{1}{\sigma_i}\Psi\left(\frac{\lambda}{\sigma_i\sqrt{-2\hat q_0}}\right)
\ee

\be\label{eq:SPConstrained2}
\Delta = \frac{1}{2\hat\Delta}\frac{1}{N}\sum_i\Phi\left(\frac{\lambda}{\sigma_i\sqrt{-2\hat q_0}}\right)
\ee

\be\label{eq:SPConstrained3}
\frac{1}{2 r} = \frac{1}{N}\sum_iW\left(\frac{\lambda}{\sigma_i\sqrt{-2\hat q_0}}\right)
\ee
and \eqref{eq:AppendixBSP1} and \eqref{eq:AppendixBSP2} remain unchanged:

\be\label{eq:SPConstrained4}
\hat\Delta=\frac{1}{2 r(1+\Delta)}
\ee

\be\label{eq:SPConstrained5}
\hat q_0=-\frac{q_0}{2 r(1+\Delta)^2}
\ee
In \eqref{eq:SPConstrained1} - \eqref{eq:SPConstrained3} we used the fact that $\Phi$, $\Psi$, and $W$ all go to zero as their argument tends to minus infinity.

Using the identity $W(x)=\frac{1}{2} x\Psi(x)+\frac{1}{2}\Phi(x)$ and the stationarity conditions above we can transform \eqref{eq:SPConstrained3} into

\be\label{eq:SPConstrained3Transf}
\lambda = \frac{q_0}{r(1+\Delta)^2},
\ee
but by \eqref{eq:SPConstrained5} this is also equal to 

\be
\lambda = -2\hat q_0.
\ee
Then the arguments of the functions $\Psi$, $\Phi$ and $W$ in \eqref{eq:SPConstrained1}--\eqref{eq:SPConstrained3} simplify as $\sqrt{\lambda}/\sigma_i$. Eq. \eqref{eq:SPConstrained3} becomes

\be\label{eq:SPConstrained3B}
\frac{1}{2r} = \frac{1}{N}\sum_i W\left(\frac{\sqrt{\lambda}}{\sigma_i}\right),
\ee 
and \eqref{eq:SPConstrained2} and \eqref{eq:SPConstrained4} combine to give for the susceptibility

\be\label{eq:SPConstrained2B}
\Delta = \frac{r\frac{1}{N}\sum_i\Phi\left(\frac{\sqrt{\lambda}}{\sigma_i}\right)}{1-r\frac{1}{N}\sum_i\Phi\left(\frac{\sqrt{\lambda}}{\sigma_i}\right)}.
\ee
Finally, for the relative estimation error, which apart form a normalizing factor is the out of sample estimator for the optimal value of risk, we find

\be\label{eq:SPConstrained4B}
q_0=\lambda r (1+\Delta)^2.
\ee
Eq. \eqref{eq:SPConstrained3B} is straightforward to solve on a machine to obtain $\lambda$ as a function of the parameters $r$, $N$, and $\sigma_i$. Once $\lambda$ is known, $\Delta$ and $q_0$ can be determined from \eqref{eq:SPConstrained2B} and \eqref{eq:SPConstrained4B}. Furthermore, by the help of the stationarity conditions we can derive the expression for the free energy

\be\label{eq:SPConstrainedFreeEnergy}
f= \frac{\lambda}{2}
\ee
 as in section 3, so the knowledge of $\lambda$ will also provide the free energy as a function of $r$, $N$, and $\sigma_i$.
 
 As for the distribution of the optimal estimated weights, by \eqref{eq:AppendixBWeightDistribution} and  \eqref{eq:AppendixBCondensate} we have
 
 \be
 p(w) = n_0\delta (w) +\theta(w)\frac{1}{N}\sum_i\frac{1}{\sigma_w^{(i)}\sqrt{2\pi}}{\rm exp}\left[-\frac{1}{2}\left(\frac{w-w_0^{(i)}}{\sigma_w^{(i)}}\right)^2\right]
 \ee
 where $\theta$ is the Heaviside function that ensures only non-negative weights appear in the distribution. The first term is the density of the weights set to zero by the no-short constraint:
 
 \be\label{eq:fractionZeroWeights}
 n_0 = \frac{1}{N}\sum_i\Phi\left(-\frac{w_0^{(i)}}{\sigma_w^{(i)}}\right).
 \ee 
The Gaussian density of the $i$-th weight is centered at $w_0^{(i)}$, which by \eqref{eq:AppendixBW1} and \eqref{eq:SPConstrained3Transf} is equal to

\be\label{eq:SPAverageWeight}
w_0^{(i)} = \frac{q_0}{(1+\Delta)}\frac{1}{\sigma_i^2},
\ee 
with standard deviation

\be\label{eq:SPStdWeight}
\sigma_w^{(i)} = \frac{\sqrt{q_0 r}}{\sigma_i}.
\ee
With this we have determined the expected positions of the estimated optimal weights and their distribution, as well as the in-sample estimated cost, and its out of sample error and susceptibility, that is we have solved the optimization of variance with a no-short-position constraint.

The limit $r\to 0 $ again corresponds to $\lambda\to\infty$, and it can easily be worked out to recover the results in Subsection 3.1, and Section 2.

\subsection{The high-dimensional regime and the critical point at $r=2$}

When $r$ is finite, we are in the high-dimensional regime where $N$ and $T$ are of the same order of magnitude. As $W$ is positive and monotonic increasing, it follows from \eqref{eq:SPConstrained3B} that with $r$ increasing $\sqrt{\lambda}$ must decrease. However, it cannot decrease below zero, and here $W(0)=1/4$, so $r$ has a maximal value $r_c=2$ beyond which it cannot grow. It seems therefore that for a given size $T$ of the samples there is an upper bound $N=2T$ beyond which we cannot consistently continue this theory. (For a physicist, all this may be vaguely reminiscent of Bose condensation.)

What is happening at $r_c=2$? First, we realize that because of the proportionality between $f$ and $\lambda$, Eq \eqref{eq:SPConstrainedFreeEnergy}, $f$ itself also has to vanish at $r=2$. But $f$ is proportional to the in-sample estimate of the portfolio variance ${\sigma^*_p}^2$, eq \eqref{eq:FreeEnergy}, so $f$ is by definition non-negative and we run into a natural bound at $r=2$.

Let us now consider the behaviour of susceptibility $\Delta$. Expanding \eqref{eq:SPConstrained3B}, \eqref{eq:SPConstrained2B} and \eqref{eq:SPConstrained4B} around $r=2$ we find 

\be
\Delta = \frac{4}{2-r},~r\to 2^-.
\ee
This reveals the meaning of the special value $r=2$: at this critical value a phase transition is taking place and the susceptibility becomes  infinitely large.
This transition may seem analogous to the one we found in the unconstrained case, but the critical value of $r$ has been shifted by the no-short constraint to $r_c=2$ from the unconstrained $r_c=1$.

There is a further difference: eq. \eqref{eq:SPConstrained4B} tells us that the behavior of $q_0$ at the phase transition is determined by the limit of $\lambda\Delta^2$ as $r\to2$.
It can be seen that

\be
\lim_{r\to 2} q_0 = \lim_{r\to 2} 2 \lambda\Delta^2 = \frac{\pi}{ \left(\frac{1}{N}\sum_i\frac{1}{\sigma_i}\right)^2}
\ee
which is finite. Therefore, in contrast to the unconstrained phase transition at $r=1$, the estimation error

\be
\tilde q_0=q_0\frac{1}{N}\sum_i\frac{1}{\sigma_i^2} = \frac{\frac{\pi}{N}\sum_i\frac{1}{\sigma_i^2}}{ \left(\frac{1}{N}\sum_i\frac{1}{\sigma_i}\right)^2} , ~r\to 2^-
\ee
remains finite. (Note that $\tilde q_0$ is larger or equal to one for any $r$, as it should, given its meaning as the relative estimation error. In particular, in the limit $r\to2$ the expression multiplying $\pi$ in the above formula is larger than equal to one for any distribution of the true variances $\sigma_i$, due to the Cauchy inequality.)

Thus the phase transition at $r=2$ displays infinite sensitivity to the input parameters, but finite estimation error.

If we picture the portfolio weights as the components of a vector then we can say that the Euclidean norm $\sum_i w_i^2$ of this vector remains finite, but the fluctuations of its direction are infinite. In other words, the longitudinal fluctuations of the weight vector have been reined in by the no-short-selling constraint, however it  was unable to suppress the transverse fluctuations. This is rather natural if we consider that the ban on short selling constrains the large compensating positions, but does not forbid the reshuffling of the components of the weight vector from sample to sample.

\begin{figure}[h]
  \centering
    \includegraphics[width=65mm]{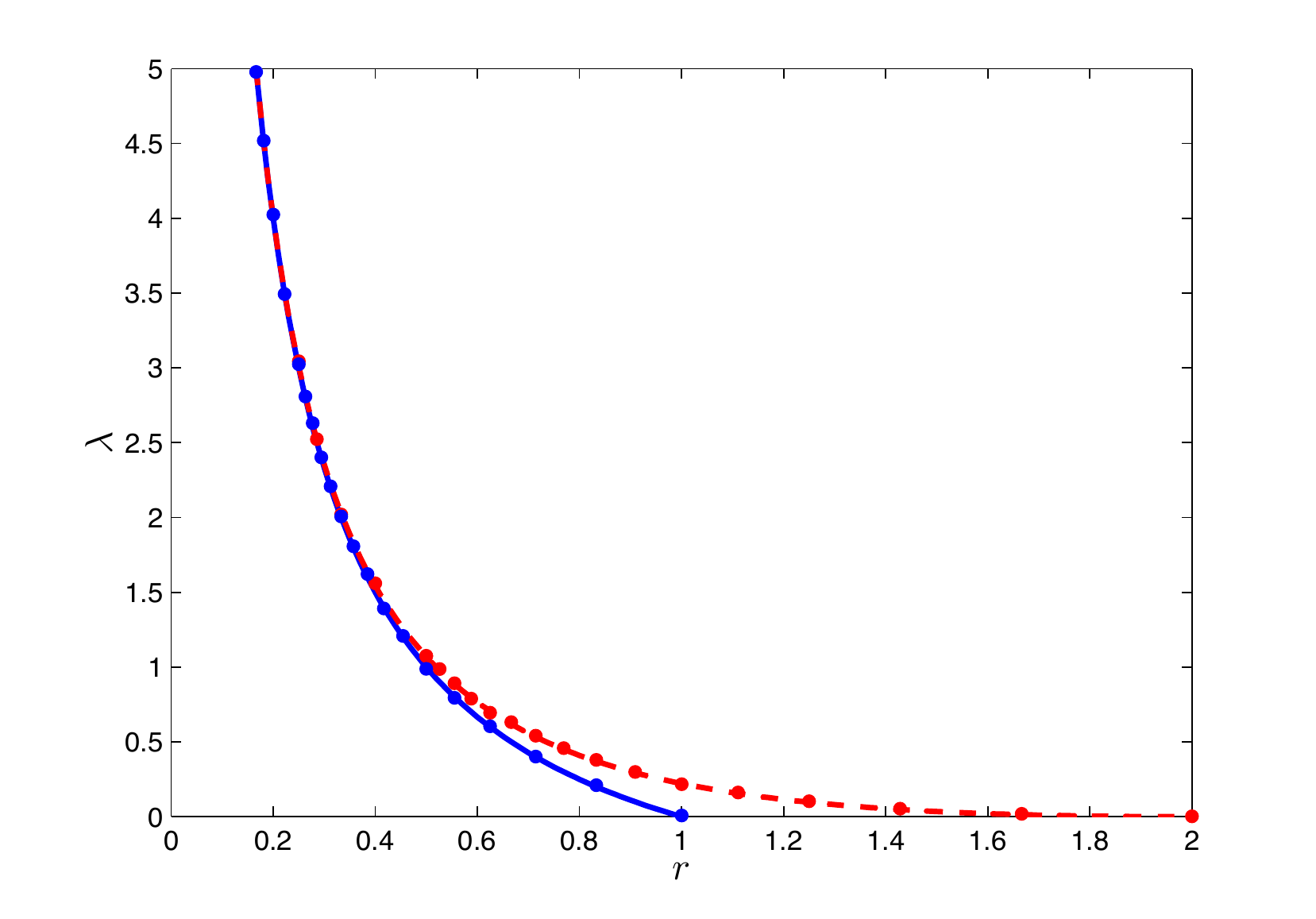}
    \includegraphics[width=65mm]{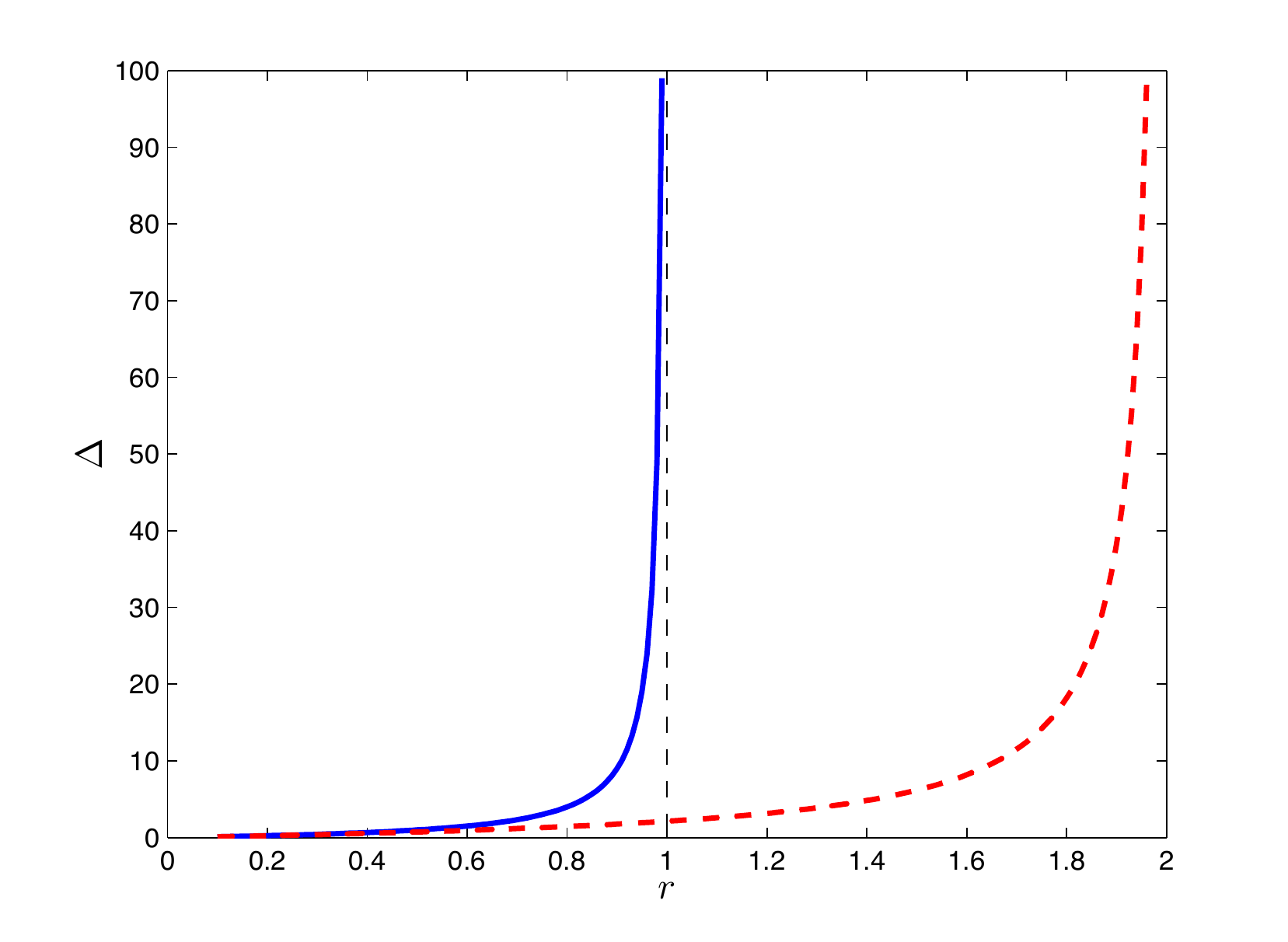}
    \includegraphics[width=65mm]{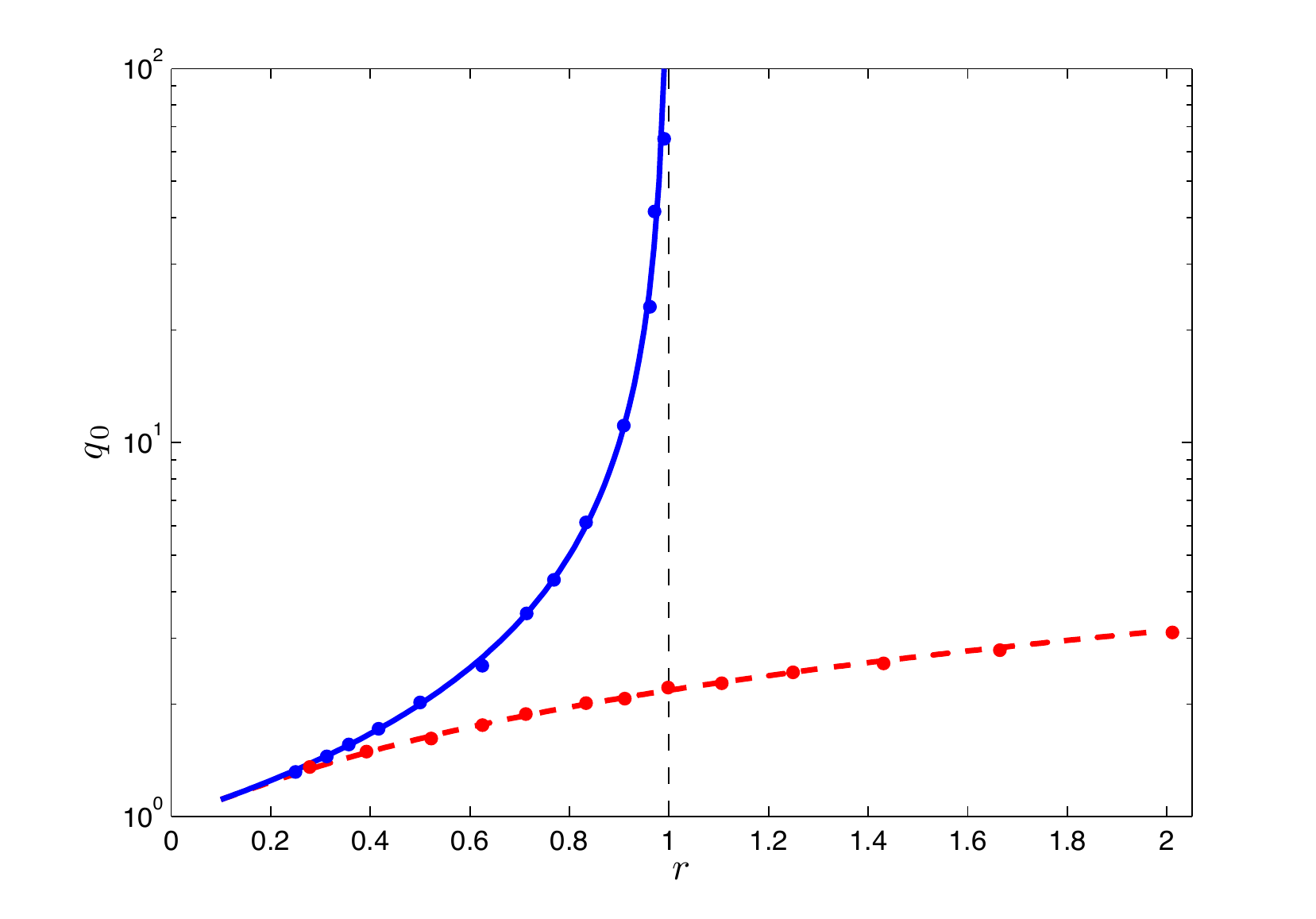}
    \caption{\footnotesize {\it The three panels show the behavior of $\lambda$ (top left panel), $\Delta$ (top right panel) and $q_0$ (bottom panel) as a function of $r$ for the cases with (solid lines) and without (dashed lines) short-selling. From the figures it is clear that the no-short selling case displays an instability at $r_c=2$. This instability is characterized by a divergent $\Delta$ (susceptibility) and a vanishing $\lambda$ (proportional to the in-sample estimate of risk), but a finite estimation error. Results of numerical simulations (dots in the top left and bottom panel) are in agreement with the analytical result.}}
\end{figure}
Fig. 2 compares the results for $\lambda$, $\Delta$ and $q_0$ as functions of $r$ for the unconstrained and the no-short-constrained cases, respectively. For simplicity, we show these results for a portfolio with all assets having the same variance $\sigma_i=1$ for all $i$.

Let us now consider the distribution of weights when $r\to 2$. Because of the divergence of $\Delta$ all the $w_0^{(i)}\to 0$, eq \eqref{eq:SPAverageWeight}. This means that the $\Phi$'s in the first term all tend to $1/2$, so the limiting density of the weights condensed at the origin becomes half of the total weight. At the same time the centers of the Gaussians in the second term will also go to zero, but according to  \eqref{eq:SPStdWeight},  their widths remain finite.

\subsection{Numerical aspects of the instability at $r=2$}

The nature of the phase transition taking place at $r=2$ is somewhat different from the one at $r=1$. While the latter takes place with probability one even for finite $N$ and $T$, the transition at $r=2$ depends on the random samples and in this respect it is rather similar to the transition in the optimization of the Expected Shortfall risk measure discussed in \cite{caccioli2015Portfolio}. Numerical experiments on small to moderate size samples of returns drawn from a symmetric distribution show that finding a solution with zero in-sample variance below $r=1$ is zero. Between $r=1$ and $r=2$ this probability is small, starting to increase as we approach $r=2$ and rapidly reaching one for $r$ values slightly exceeding 2. The transition is the faster the larger the dimension and becomes sharp in the limit of high dimensions. The zero in-sample variance solutions above $r=2$ are the natural continuations of the analytic result for the vanishing in-sample estimator at $r=2$ and also share its infinite susceptibility. Thus, in fact, in high dimensions and above the critical ratio $r=2$, we again have a continuum of solutions to the optimization problem, corresponding to a flat cost landscape for each sample. These solutions are infinitely sensitive to any change in the input data, and jump about the landscape from sample to sample, in accord with the infinite susceptibility. 

Concerning numerical work, it perhaps requires even more care now than around the $r=1$ phase transition. At variance with that, the instability at $r=2$ is not accompanied by large fluctuations in the estimated cost, it is more subtle, it corresponds to the fluctuations of the direction of the weight vector. Some of the standard solvers do signal the problem when encountering a singular covariance matrix, others take care of the difficulty by regularizing the problem on their own. It is the obligation of the user to carefully acquaint herself with the details of the solver instead of accepting a seemingly stable answer to a meaningless question.

\subsection{Preferential elimination of large volatility assets}

The constraint on short positions is a special case of $\ell_1$ regularization. As such, it is expected to result in a sparse optimal portfolio, that is to eliminate some of the assets. The build-up of the weight at $w=0$ is the consequence if this tendency of $\ell_1$.

Our results do not refer to a single sample, but to averages over the samples. On average, each asset contributes to the peak of the weight distribution at $w=0$, i.e. each asset gets eliminated with a certain probability. However, the probability of getting eliminated depends on the variance of the given asset.

The argument of the function $\Phi$ in \eqref{eq:fractionZeroWeights} is

\be\label{eq:420}
-\frac{w_0^{(i)}}{\sigma_w^{(i)}} = - \left(\frac{q_0}{r}\right)^{1/2}\frac{1}{1+\Delta}\frac{1}{\sigma_i}
\ee
and $\Phi$  is monotonic increasing. Accordingly, assets with a large standard deviation (large volatility) become eliminated with larger probability than those with a small volatility. This selection is particularly strong when the coefficient of $1/\sigma_i$ in \eqref{eq:420} is large, that is $r$ is small, while the distinction between high and low volatility items disappears as we approach $r=2$, where $\Delta\to\infty$. This is plausible: if we have a lot of information ($r$ small) the regularizer can clearly distinguish between the low and high volatility items, but when fluctuations dominate any possibility of making a difference vanishes.

Note that as $\Phi(0)=1/2$ and the argument of $\Phi$ in \eqref{eq:fractionZeroWeights} is always negative, in the limit $r\to 2$ the ``condensate density'' $n_0$ approaches its maximal value $1/2$ from below: the no-short constraint pushes at most half of the assets into the ``condensate''. At the same time, because of the divergence of $\Delta$ the centers of the Gaussians also shift to the origin, but their standard deviations remain finite.

\section{Summary}
Let us briefly summarize the main results of this paper. We have considered a portfolio in the high-dimensional limit where the number of different assets $N$ and the sample size $T$ are large, with their ratio $r=N/T$ kept finite. We assumed that the returns on the assets were independent normal variables with zero expected value and different variances. We optimized the variance of the portfolio under the budget constraint with or without a ban on short positions and averaged the results over the random samples by the method of replicas borrowed from the statistical physics of disordered systems. 

In the simple case where unlimited short positions were allowed we recovered known results for the out of sample estimator for the variance, the susceptibility (sensitivity of the estimator to small variations in the input data), and the in-sample average of the portfolio variance. As a new result, we also derived the distribution of optimal weights over the random samples. We found that the originally sharply distinguishable spikes of this distribution broaden with increasing $r$ until in the limit $r\to 1$ any distinction between the different weights gets completely washed away due to the divergent sample fluctuations. In the same limit the estimation error and the susceptibility diverge and the in-sample variance of the portfolio vanishes; at $r=1$ a phase transition is taking place. This is the same point where the first zero eigenvalue of the covariance matrix appears. Beyond this critical value of $r$ the variance cannot be meaningfully optimized: a continuum of solutions appear, since any combination of the zero eigenvectors of the covariance matrix make the variance zero. As argued above, the phenomenon does not depend on the use of the replica method or the assumption about the Gaussian distribution of returns: it is a purely geometric effect, depending solely on the fact that the rank of the covariance matrix is the smaller of $N,T$ in any sample, with probability one.  

In order to support and illustrate the theoretical results, we also solved the quadratic programming task of optimizing the variance numerically. While the agreement between the analytic theory and numerics is perfect below the critical point $r=1$, for $r>1$ we found that some standard solvers continue to find a stable, unique solution with all the optimal weights the same, the portfolio variance identically zero and the estimation error and susceptibility decreasing with $r$ increasing further. This apparent restabilization is an artifact, due to a built-in stabilizing feature (essentially an $\ell_2$ regularizer) in the solvers.

The main result of the paper is the solution of variance optimization under a ban on short positions. This problem, which has a great importance in practice, had not been solved analytically before. The method of replicas allowed us to derive results for the same quantities as in the previous case: we have determined the out of sample estimator for the variance, the susceptibility, the optimal in-sample variance and the distribution of optimal portfolio weights again. The constraint on short positions acts as a kind of $\ell_1$ regularizer and eliminates some (at most half) of the assets resulting in a sparser portfolio. Accordingly, a sharp peak is built up in the distribution of weights at the origin and the remaining weights are all positive.  In agreement with one's natural expectation, assets with larger volatility get eliminated with higher probability than the low volatility items.

It might have been expected that the constraint on short positions would tame the large sample fluctuations. This expectation is borne out only partially: it is true that the optimization can now be performed also above the previous critical value $r=1$, but at $r=2$ we discover another phase transition. This time the estimation error stays finite, but the susceptibility still diverges here. The in-sample estimator for the portfolio variance vanishes, and the distribution of weights is smeared out again. 

Numerical work on finite size samples shows that the probability of solutions with zero in-sample variance is zero below $r=1$, very small between $r=1$ and $r=2$, and rapidly goes to one above $r=2$. Accordingly, we find perfect agreement between the analytic theory and simulations below the $r=2$ transition already for moderate sized samples, but in the region above $r=2$, where the instability prevents the analytic theory to penetrate, a continuum of unstable, zero-variance solutions arises, with a flat cost-landscape. Concerning the application of solvers from libraries such as R or Matlab, our experience is similar to that around the $r=1$ transition. Some standard solvers keep finding a stable, unique solution with all the weights the same also above $r=2$ where we know that a continuum of solutions exist, and the solvers should, in principle, obtain an unstable solution, different in each sample. The explanation of the phenomenon is the same as in the unconstrained case: these solvers are built in such a way as to find the diagonal solution whenever the covariance matrix has zero modes.

The financial content of the instabilities described above is the following. When unlimited short positions are allowed one can assume very large compensating positive and negative positions without violating the budget constraint and keeping the portfolio variance low. As the dimension of the portfolio increases, the (Euclidean) length of the weight vector, hence also the leverage, diverge -- a fundamentally risky situation around a point where the estimated portfolio variance vanishes. When we switch on the constraint on short positions, it becomes impossible to build up large compensating positions, and the length of the weight vector, hence also the estimation error, remain finite, but the solution is still unstable with respect to rearrangements, or simply to a reshuffling of the components of the optimal weight vector from sample to sample. This corresponds to  divergent transverse fluctuations of the weight vector, indicated by the divergent susceptibility.

A final remark on the miraculous restabilization of the numerical solutions: in empirical work where one has real life data without the luxury of a large number of samples to average over, one may easily overlook the instability in the no-short-selling case, especially if the software package is a black box for the portfolio manager. We think one should never use a ready-made program without the detailed knowledge of the algorithm implemented in it. Furthermore, one should never trust a purely numerical result without an understanding of the main structural features of the problem, such as the instability described here. Although seldom able to follow it, we agree with Lev Landau's maxim: one should not attempt to solve a problem before knowing the solution in advance.

\appendix
\section{Derivation of the free energy with the replica method}

We consider the following problem: given a financial market where $N$ risky assets are traded we want to find the portfolio $\vec w$ that minimizes the risk function
\be\label{markowitz}
R(\vec w)=\frac{1}{2}\sum_{i,j}w_iC_{ij}w_j,
\ee
under the budget constraint $\sum_{i=1}^N w_i=N$.
In the above expression $w_i$ represents the position held on asset $i$, while $C_{ij}$ is the assets covariance matrix.
In practice, the true covariance matrix is unknown and one has to rely on estimators based on historical data. If $x_{it}$ represents the return of asset $i$ at time $t$, the entries of the covariance matrix can be estimated as 
\be
C_{ij}=\frac{1}{T}\sum_{t=1}^T x_{it}x_{jt}.
\ee
Furthermore, we consider adding the following term (an asymmetric $\ell_1$ regularizer) to the cost function 
\be
g(\vec w) = \eta_1 \sum_i w_i \theta(w_i) - \eta_2 \sum_i w_i \theta(-w_i),
\ee
so that the optimization problem becomes 
\bea
&{\rm min}_{\vec w} &\big\{\frac{1}{2} \sum_{ij}w_i x_{it}x_{jt}w_j+g(\vec w) \big\}\\
&{\rm s.t.}& \sum_i w_i=N,
\eea
where for later convenience we have multiplied the empirical covariance matrix by a factor $T$.
In the following we assume that the $x_{it}$ are drawn from independent Gaussian distributions of zero mean and variance $\sigma_i^2/N$.\\ 
Taking advantage of the identity
\be
\langle (\log Z)^n\rangle = \Big\langle\frac{Z^n-1}{n}\Big\rangle,
\ee
valid in the limit $n\to0$, the typical properties of the solution can be captured by computing the replicated partition function
\be
Z_n(\vec w)=\Big\langle \int_{-\infty}^\infty \prod_{i=1}^N\prod_{a=1}^n dw_i^a e^{-\gamma\left(\frac{1}{2}\sum_{i,j,t,a}w_i^a x_{it}x_{jt}w_j^a+g(\w)\right)}\prod_a\delta(\sum_i w_i^a-N)\Big\rangle_{\x_{t}}
\ee
and then taking the limits
\be
\lim_{\gamma\to\infty}\lim_{n\to0} \frac{1}{\gamma} Z_n(\vec w),
\ee
where $\gamma$ is a fictitious inverse temperature that we introduce to simplify the calculation and $\langle\cdots\rangle$ represents an average over the probability distribution of returns.
The above partition function refers to a system of $n$ replicas of the original system,
and  the index $a$ is introduced  to label different replicas, so that $w_i^a$ represents the $i$-th weight of the $a$-th replica.  
Introducing an integral representation for the delta function and performing a Hubbard-Stratonovich transformation the replicated partition function can be written as
\beas
Z_n(\vec w)&=&\Big\langle \int_{-\infty}^\infty \prod_{i,a,t}^N dw_i^a d\phi_{at} d\lambda^a{\rm exp}\left[-\frac{1}{2}\sum_{a,t}\phi_{at}^2+i\sqrt{\gamma}\sum_{i,t,a}\phi_t^a w_i^a x_{it}\right]\\
&\times&\exp \left[\sum_a\lambda^a(\sum_i w_i^a-N)-\gamma g(\w) \right]\Big\rangle_{\x_{t}}.
\eeas
Averaging over the probability distributions of returns gives
\beas
Z_n(\vec w)&=& \int_{-\infty}^\infty \prod_{i,a,b,t} dw_i^a d\hat{Q}_{ab} d\phi_{at} d\lambda^a \exp\left[-\frac{1}{2}\sum_{a,t}\phi_{at}^2-\frac{\gamma}{2}\sum_{a,b,t}\phi_{at}Q_{ab}\phi_{b,t}\right]\\
&\times&\exp\left[\sum_{a,b}\hat{Q}_{ab}\left(NQ_{ab}-\sum_i \sigma_i^2 w_i^aw_i^b\right)+\sum_a\lambda^a\left(\sum_i w_i^a-N\right)-\gamma g(\w)\right]\\
\eeas
where we have introduced the overlap matrix $Q_{ab}=\frac{1}{N}\sum_i\sigma_i^2 w_i^aw_i^b$ and the conjugate variables $\hat{Q}_{ab}$ to enforce this relation.\\
We can now integrate over the variables $\phi_{at}$ to obtain
\beas
Z_n(\vec w)&=& \int_{-\infty}^\infty \prod_{i,a,b,t} dw_i^a d\hat{Q}_{ab} d\lambda^a\exp\left[-\frac{T}{2}{\rm  tr}\log\left(\delta_{ab}+\gamma Q_{ab}\right)\right]\\ 
&\times&\exp\left[\sum_{a,b}\hat{Q}_{ab}\left(NQ_{ab}-\sum_i \sigma_i^2 w_i^aw_i^b\right)+\sum_a\lambda^a\left(\sum_i w_i^a-N\right)-\gamma g(\w) \right]\\
\eeas
The convexity of the cost function motivates the choice of the replica symmetric ansatz
\begin{equation}
    Q_{ab}= \left\{ \begin{array}{cc} q_0+\Delta ,&    a =b\\
    q_0 , &  a\neq b \end{array} \right.
\end{equation}
\begin{equation}
    \hat{Q}_{ab}= \left\{ \begin{array}{cc} \hat{q}_0+\hat{\Delta} ,&    a = b  \\
    \hat{q}_0 , &  a\neq b . \end{array} \right.
\end{equation}
To leading order in $n$ we have 
\bea
-\frac{T}{2}{\rm tr}\log(\delta_{ab}+\gamma Q_{ab})&=&-\frac{T}{2}\left[\log\left(1+\gamma\Delta\right)+\frac{\gamma q_0}{1+\gamma\Delta}\right]\\
\sum_{a,b}\hat{Q}_{ab}Q_{ab}&=&Nn(\hat{q}_0\Delta+q_0\hat{\Delta}+\Delta\hat{\Delta}),
\eea
while the $\w$-dependent part of the partition function can be written as
\be
\int  d\lambda^a d\hat{\Delta}d\hat{q}_0\exp\left[Nn\Big\langle \log\int dw e^{-\hat{\Delta}\sigma^2 w^2+w z \sigma \sqrt{-2\hat{q}_0}+\lambda w- g(\w)]}\Big\rangle_{z\sigma}\right],
\ee
where $\langle \cdots \rangle_{z\sigma}$ denotes averages over the normal variable $z$ and the distribution of asset variances:
\be
\langle h(z,\sigma) \rangle_{z\sigma} = \int d\sigma \frac{1}{N}\sum_i  \delta(\sigma-\sigma_i) \left(\int_{-\infty}^\infty \frac{dz}{\sqrt{2\pi}}h(z,\sigma) e^{-z^2/2}\right).
\ee
If we now write the partition function as
\be\label{partfin}
Z_n=\int d\lambda dq_0d\Delta d\hat{q}_0d\hat{\Delta}e^{-\gamma nN f(\lambda,q_0,\Delta,\hat{q}_0,\hat{	\Delta})},
\ee
we find
\beas
 f(\lambda,q_0,\Delta,\hat{q}_0,\hat{	\Delta})&=&\frac{1}{2\gamma r}\left[\log(1+\gamma\Delta)+\frac{\gamma q_0}{1+\gamma\Delta}\right]+\frac{\lambda}{\gamma}
 -\frac{1}{\gamma}(\hat{q}_0\Delta+q_0\hat{\Delta}+\Delta\hat{\Delta})\\
\nonumber &-&\frac{1}{\gamma}\Big\langle \log \int dw e^{-\hat{\Delta}\sigma^2 w^2+w z\sigma \sqrt{-2\hat{q}_0}+\lambda w - g(\w)} \Big\rangle_{z\sigma}
\eeas
Performing the change of variables $\Delta\to\Delta/\gamma$, $\hat{q}_0\to\gamma^2\hat{q}_0$, $\hat{\Delta}\to\gamma\hat{\Delta}$, $\lambda\to\gamma\lambda$ and taking the limit $\gamma\to\infty$ we finally have

\be
 f(\lambda,q_0,\Delta,\hat{q}_0,\hat{\Delta}) = \frac{q_0}{2 r (1+\Delta)}-\hat{q}_0\Delta-\hat{\Delta} q_0+\lambda+ {\min_{\w}} \Big\langle V(\w)  \Big\rangle_{z\sigma},
\ee
where
 \be
 V = \hat{\Delta} \sigma^2 w^2-w z \sigma \sqrt{-2\hat{q}_0}-\lambda w + \eta_1\theta(w)-\eta_2\theta (-w).
 \ee

\section{The saddle point conditions and the distribution of weights}

In Appendix A we derived the free energy functional

\be\label{eq:AppendixBFreeEnergy}
 f(\lambda,q_0,\Delta,\hat{q}_0,\hat{\Delta}) = \frac{q_0}{2 r (1+\Delta)}-\hat{q}_0\Delta-\hat{\Delta} q_0+\lambda+ {\min_{\w}} \Big\langle V(\w)  \Big\rangle_{z\sigma},
\ee
where the ``potential'' is

 \be\label{eq:AppendixBPotential}
 V = \hat{\Delta} \sigma^2 w^2-w z \sigma \sqrt{-2\hat{q}_0}-\lambda w + \eta_1\theta(w)-\eta_2\theta (-w).
 \ee
 The double averaging $\langle \dots \rangle_{\sigma,z}$ means

\be\label{eq:AppendixBDoubleAverage}
\int_0^\infty d\sigma\frac{1}{N} \sum_i \delta(\sigma-\sigma_i)\int_{-\infty}^\infty\frac{dz}{\sqrt{2\pi}}e^{-z^2/2}\ldots
\ee
The potential does not contain $q_0$ and $\Delta$, therefore the saddle point (or stationarity) conditions can be written up for these variables immediately

\be\label{eq:AppendixBSP1}
\frac{\partial f}{\partial q_0}=0 \Rightarrow \hat\Delta = \frac{1}{2r(1+\Delta)},
\ee

\be\label{eq:AppendixBSP2}
\frac{\partial f}{\partial \Delta}=0 \Rightarrow \hat q_0 = - \frac{q_0}{2r(1+\Delta)^2}.
\ee
From these the useful combination

\be\label{eq:AppendixBUsefulCombination}
\sigma_w = \frac{\sqrt{-2\hat q_0}}{2\hat \Delta} = \sqrt{q_0 r}
\ee
can be obtained.

Here and in the following we will frequently encounter the integrals of the standard normal distribution:

\be
\nonumber\Phi(x) = \int_{-\infty}^ x\frac{dt}{\sqrt{2\pi}} e^{-t^2/2},
\ee

\be
\nonumber\Psi(x) = \int_{-\infty}^ x dt \Phi(t),
\ee

\be
\nonumber W(x) =  \int_{-\infty}^ x dt \Psi(t).
\ee

The minimum of the potential is at

\be
w^* = \frac{\sigma z \sqrt{-2\hat q_0}+\lambda-\eta_1\theta(w^*)+\eta_2\theta(-w^*)}{2\hat\Delta\sigma^2}.
\ee

Substituting this back into \eqref{eq:AppendixBPotential} and performing the double average according to the recipe in \eqref{eq:AppendixBDoubleAverage} we find that the last term in \eqref{eq:AppendixBFreeEnergy} is

\be
\langle V^*\rangle_{ z\sigma} = \frac{\hat q_0}{\hat\Delta}\frac{1}{N}\sum_i \left( W\left(\frac{\lambda-\eta_1}{\sigma_i\sqrt{-2\hat q_0}}\right) + W\left(-\frac{\lambda +\eta_2}{\sigma_i\sqrt{-2\hat q_0}}\right)\right).
\ee
Then the free energy becomes

\be\label{eq:AppendixBFreeEnergy2}
f = \lambda-\Delta\hat q_0-\hat\Delta q_0+\frac{q_0}{2r(1+\Delta)}+\frac{\hat q_0}{\hat\Delta}\frac{1}{N} \sum_i \left( W\left(\frac{\lambda-\eta_1}{\sigma_i\sqrt{-2\hat q_0}}\right) + W\left(-\frac{\lambda +\eta_2}{\sigma_i\sqrt{-2\hat q_0}}\right)\right)
\ee
The remaining three saddle point equations are obtained by taking the derivatives of the above expression with respect to $\lambda$, $\hat\Delta$ and $\hat q_0$ respectively.

\be
\nonumber \frac{\partial f}{\partial\lambda} = 0\Rightarrow 1+\frac{\hat q_0}{\hat\Delta} \frac{1}{N} \sum_i\frac{1}{\sigma_i\sqrt{-2\hat q_0}} \left( \Psi\left(\frac{\lambda-\eta_1}{\sigma_i\sqrt{-2\hat q_0}}\right) - \Psi\left(-\frac{\lambda +\eta_2}{\sigma_i\sqrt{-2\hat q_0}}\right)\right) = 0
\ee
or, with \eqref{eq:AppendixBUsefulCombination},

\be\label{eq:AppendixBSP3}
\frac{1}{\sqrt{q_0 r}} =  \frac{1}{N} \sum_i\frac{1}{\sigma_i} \left( \Psi\left(\frac{w_1^{(i)}}{\sigma_w^{(i)}}\right) - \Psi\left(-\frac{w_2^{(i)}}{\sigma_w^{(i)}}\right)\right). 
\ee
Here the notations
\be\label{eq:AppendixBW1}
w_1^{(i)} = \frac{\lambda-\eta_1}{2\sigma_i^2 \hat \Delta}=\frac{(\lambda-\eta_1)r (1+\Delta)}{\sigma_i^2},
\ee
\be\label{eq:AppendixBW2}
w_2^{(i)} = \frac{\lambda+\eta_2}{2\sigma_i^2 \hat \Delta}=\frac{(\lambda+\eta_2)r (1+\Delta)}{\sigma_i^2}
\ee
and
\be\label{eq:AppendixBSigma}
\sigma_w^{(i)} = \frac{\sigma_w}{\sigma_i} = \frac{\sqrt{q_0 r}}{\sigma_i}
\ee
have been introduced.

\be\label{eq:AppendixBSP4}
\frac{\partial f}{\partial\hat q_0} = 0\Rightarrow  \Delta = \frac{1}{2\hat\Delta N} \sum_i \left( \Phi\left(\frac{w_1^{(i)}}{\sigma_w^{(i)}}\right) + \Phi\left(-\frac{w_2^{(i)}}{\sigma_w^{(i)}}\right)\right). 
\ee
where the identity $W(x)=\frac{1}{2}x\Psi(x)+\frac{1}{2}\Phi(x)$ has been used.
With \eqref{eq:AppendixBSP1} we can cast \eqref{eq:AppendixBSP4} into the form

\be
\Delta = \frac{\frac{r}{N}\sum_i\left(\Phi\left(\frac{w_1^{(i)}}{\sigma_w^{(i)}}\right)+\Phi\left(-\frac{w_2^{(i)}}{\sigma_w^{(i)}}\right)\right)}{1-\frac{r}{N}\sum_i\left(\Phi\left(\frac{w_1^{(i)}}{\sigma_w^{(i)}}\right)+\Phi\left(-\frac{w_2^{(i)}}{\sigma_w^{(i)}}\right)\right)}.
\ee
Finally

\be
\nonumber\frac{\partial f}{\partial \hat\Delta} = 0\Rightarrow q_0=-\frac{\hat q_0}{\hat\Delta^2}\frac{1}{N}\sum_i \left(W\left(\frac{w_1^{(i)}}{\sigma_w^{(i)}}\right)+W\left(-\frac{w_2^{(i)}}{\sigma_w^{(i)}}\right)\right),
\ee
which can be written by help of \eqref{eq:AppendixBSP1}, \eqref{eq:AppendixBSP2} as

\be\label{eq:AppendixBSP5}
\frac{1}{2r} = \frac{1}{N}\sum_i \left(W\left(\frac{w_1^{(i)}}{\sigma_w^{(i)}}\right)+W\left(-\frac{w_2^{(i)}}{\sigma_w^{(i)}}\right)\right).
\ee
The distribution of weights can be obtained from

\be
\nonumber p(w)=\langle \delta(w-w^*)\rangle_{z\sigma}
\ee
and works out to be

\bea
\nonumber p(w) &=& \frac{1}{N}\sum_i \left(\Phi\left(\frac{-w_1^{(i)}}{\sigma_w^{(i)}}\right)-\Phi\left(-\frac{w_2^{(i)}}{\sigma_w^{(i)}}\right)\right)\delta(w)\\
\nonumber &+& \frac{1}{N}\sum_i \frac{1}{\sigma_w^{(i)}\sqrt{2\pi}}{\rm exp}\left(-\frac{1}{2}\left(\frac{w-w_1^{(i)}}{\sigma_w^{(i)}}\right)^2\right)\theta(w)\\
&+& \frac{1}{N}\sum_i  \frac{1}{\sigma_w^{(i)}\sqrt{2\pi}}{\rm exp}\left(-\frac{1}{2}\left(\frac{w-w_2^{(i)}}{\sigma_w^{(i)}}\right)^2\right)\theta(-w)\label{eq:AppendixBWeightDistribution}
\eea
Here, the first term is the density of the zero weights

\be\label{eq:AppendixBCondensate}
n_0\equiv \frac{1}{N}\sum_i \left(\Phi\left(\frac{w_2^{(i)}}{\sigma_w^{(i)}}\right)-\Phi\left(\frac{w_1^{(i)}}{\sigma_w^{(i)}}\right)\right),
\ee
and

\be
n_0^{(i)} = \frac{1}{N} \left(\Phi\left(\frac{w_2^{(i)}}{\sigma_w^{(i)}}\right)-\Phi\left(\frac{w_1^{(i)}}{\sigma_w^{(i)}}\right)\right),
\ee
is the contribution of the $i$-th asset to this ``condensate''. The appearance of this term is due to the $\ell_1$ regularizer.

The distribution of the non-zero weights is given by the second and third terms of \eqref{eq:AppendixBWeightDistribution}. This formula reveals the meaning of the symbols introduced in \eqref{eq:AppendixBW1}, \eqref{eq:AppendixBW2} and \eqref{eq:AppendixBSigma}: $w_1^{(i)}$ and $w_2^{(i)}$ are the centers of the two Gaussians in \eqref{eq:AppendixBWeightDistribution}, while $\sigma_w^{(i)}$ their standard deviation.

\section*{Acknowledgements}

Valuable discussions with Risi Kondor and Istv\'an Csabai are greatly appreciated. We are also obliged to Barbara D\"om\"ot\"or for calling Jorion's paper to our attention, to A. Eggington for reading the manuscript, and to Jo\" el Bun for useful comments. F.C. acknowledges support of the Economic and Social Research Council (ESRC) in funding the Systemic Risk Centre (ES/K002309/1).
\bibliographystyle{unsrt}
\bibliography{noShort}
\end{document}